\documentclass{icrc2009}

\usepackage{graphicx}   
\usepackage{caption}    
\usepackage{subfig}
\usepackage{fixltx2e}
\usepackage{url}

\newcommand{\shorttitle}[1]%
{\markboth{Proceedings of the 31\MakeLowercase{$^{st}$} ICRC, {\L}\'{o}d\'{z} 2009}{#1} }
\newcommand{\etal}{\MakeLowercase{\textit{et al. }}} 

\begin{document}
\title{Measurement of the antiproton/proton ratio at few-TeV energies with the ARGO-YBJ experiment}

\author{\IEEEauthorblockN{Giuseppe Di Sciascio\IEEEauthorrefmark{1},
                           Roberto Iuppa\IEEEauthorrefmark{1}
                           \IEEEauthorrefmark{2} and
                           Silvia Vernetto\IEEEauthorrefmark{3} \IEEEauthorrefmark{4}\\
                           on behalf of the ARGO-YBJ Collaboration}
                            \\
\IEEEauthorblockA{\IEEEauthorrefmark{1} INFN, Sezione Roma Tor
Vergata, via della Ricerca Scientifica 1, Rome - Italy}
 \IEEEauthorblockA{\IEEEauthorrefmark{2} Dipartimento di Fisica,
 Universit\'a Roma Tor Vergata, via della Ricerca Scientifica 1, Rome - Italy}
  \IEEEauthorblockA{\IEEEauthorrefmark{3} IFSI-INAF, Corso Fiume 4, Torino -
  Italy}
  \IEEEauthorblockA{\IEEEauthorrefmark{4} INFN Sezione di Torino, Via Pietro Giuria 1, Torino - Italy}}

\shorttitle{G. Di Sciascio \etal $\bar{p}/p$ ratio measurement}
\maketitle

\begin{abstract}
Cosmic ray antiprotons provide an important probe for the study of
cosmic-ray propagation in the interstellar space and to
investigate the existence of Galactic dark matter. Cosmic rays are
hampered by the Moon, therefore a deficit of cosmic rays in its
direction is expected (the so-called \emph{Moon shadow}). The
Earth-Moon system acts as a magnetic spectrometer. In fact, due to
the geomagnetic field the center of the Moon shifts westward by an
amount depending on the primary cosmic ray energy. Paths of
primary antiprotons are therefore deflected in an opposite sense
in their way to the Earth. This effect allows, in principle, the
search of antiparticles in the opposite direction of the observed
Moon shadow.

The ARGO-YBJ experiment, in stable data taking since November 2007
with an energy threshold of a few hundreds of GeV, is observing
the Moon shadow with high statistical significance. Using about 1
year data, an upper limit of the $\bar{p}/p$ flux ratio in the
few-TeV energy region is set to a few percent with a confidence
level of 90\%.
  \end{abstract}

\begin{IEEEkeywords}
 ARGO-YBJ experiment, Antiproton/proton ratio measurement, TeV Dark Matter
\end{IEEEkeywords}

\section{Introduction}

The study of cosmic ray (CR) antiprotons may present an
opportunity to investigate the baryonic asymmetry of the universe
and to uncover evidence for the existence of galactic dark matter.
Antiprotons are anyway produced by standard nuclear interactions
of CR nuclei over the interstellar medium (spallation processes).
Therefore, the observation of $\overline{p}$ abundance in the CR
flux is a key to understanding CR propagation and provide
information complementary to that provided by secondary nuclei as
Li, Be and B or secondaries of iron. Unlike secondary nuclei,
antiprotons are tracers primarily of the propagation history of
the proton, the dominant CR component \cite{strong07}. Recent
measurements of the antiproton flux up to about 100 GeV
\cite{bess,caprice,pamelap} are consistent with the conventional
CR model, in which antiprotons are secondary particles yielded by
the spallation of CR nuclei over the interstellar medium.

Nevertheless, exotic models of primary $\overline{p}$ production
not ruled out by low energy measurements yet are available.
Antiprotons can be produced from primordial black hole evaporation
or in anti-galaxies. In some scenarios models of primary
antiprotons production can provide a $\overline{p}/p$ ratio
increasing up to the 10\% level in the few-TeV energy range (see
the review paper by Stephen and Golden \cite{golden}).

In addition, cosmic ray antiprotons, as well as positrons, are
considered as prime targets for indirect detection of galactic
dark matter (see for example \cite{donato} and reference therein).
As an example, some recent analyses suggest that the overall
PAMELA \cite{pamelap,pamelae} $\overline{p}$ and $e^+$ data and
ATIC/PPB-BETS \cite{atic,ppb} $e^++e^-$ data can be reproduced
taking into account a heavy DM particle (M $\geq$10 TeV) that
annihilates into $W^+W^-$ or $hh$ \cite{cirelli}. This scenario
implies that the $\overline{p}/p$ ratio, consistent with the
background of secondary production up to about 50 GeV, increases
rapidly reaching the 10$^{-2}$ level in the TeV energy region.

Linsley \cite{linsley} and Lloyd-Evans \cite{lloyd} in 1985
independently explored the possibility to use the Moon or Sun
shadows as mass spectrometer in order to measure the charge
composition of cosmic ray spectrum. In particular Linsley first
discussed the idea to measure the cosmic ray antiprotons abundance
exploiting the separation of the proton and antiproton shadows.
The geomagnetic field should deflect the antimatter component of
the cosmic rays in the opposite direction with respect to the
matter one. Therefore, if protons are deflected by the geomagnetic
field towards east, antiprotons are deflected towards west. If the
energy and the angular resolution are respectively low and small
enough, we can distinguish, in principle, between two shadows, one
shifted towards west due to the protons and the other shifted
towards east due to the antiprotons one. At high energy ($\geq$ 10
TeV) the magnetic deflection is too small compared to the angular
resolution and the shadows cannot be disentangled. At low energy
($\approx$100 GeV) the well separated shadows are washed out by
the poor angular resolution, thus limiting the sensitivity.
Therefore, there is an optimal energy window for the measurement
of the antiproton abundance.

In this paper we report on the measurement of the $\overline{p}/p$
ratio in the few-TeV energy region exploiting the observation of
the galactic cosmic ray Moon shadowing effect performed by the
ARGO-YBJ experiment.

\section{The ARGO-YBJ experiment}

The ARGO-YBJ detector, located at the YangBaJing Cosmic Ray
Laboratory (Tibet, P.R. China, 4300 m a.s.l.), is the only
experiment exploiting the \emph{full coverage} approach at very
high altitude. The detector is constituted by a central carpet
$\sim$74$\times$78 m$^2$, made of a single layer of Resistive
Plate Chambers (RPCs) with $\sim$92$\%$ of active area, enclosed
by a partially instrumented guard ring that extends the detector
surface up to $\sim$100$\times$110 m$^2$. The apparatus has a
modular structure, the basic data acquisition element being a
cluster (5.72$\times$7.64 m$^2$), namely a group of 12 RPCs
(2.80$\times$1.25 m$^2$ each). Each chamber is read by 80 strips
of 7$\times$62 cm$^2$ (the spatial pixel), logically organized in
10 independent pads of 56$\times$62 cm$^2$ representing the time
pixel of the detector. The RPCs are operated in streamer mode with
a standard gas mixture (Argon 15\%, Isobutane 10\%,
TetraFluoroEthane 75\%), the High Voltage settled at 7.2 kV
ensures an overall efficiency of about 96\% \cite{aielli06}. The
central carpet contains 130 clusters (hereafter ARGO-130) and the
full detector is composed of 153 clusters for a total active
surface of $\sim$6700 m$^2$.

All events giving a number of fired pads N$_{pad}\ge$ N$_{trig}$
in the central carpet within a time window of 420 ns are recorded.
The spatial coordinates and the time of any fired pad are then
used to reconstruct the position of the shower core and the
arrival direction of the primary.

The ARGO-YBJ experiment started recording data with the whole
central carpet in June 2006. The period until autumn 2007 has been
mainly devoted to installation and debugging operations, the duty
cycle being lower in that period. Since 2007 November the full
detector is in stable data taking at the multiplicity trigger
threshold N$_{trig}\geq$20 and a duty cycle $\sim 90\%$: the
trigger rate is about 3.6 kHz.

The reconstruction of shower parameters is split into the
following steps. First the shower core position is derived with
the Maximum Likelihood method from the lateral density
distribution of the secondary particles. In the second step, given
the shower core position, the shower axis is reconstructed by
means of an iterative un-weighted planar fit being able to reject
the time values belonging to the non-gaussian tails of the arrival
time distributions. A conical correction with a slope fixed to
$\alpha$ = 0.03 rad is applied to the surviving hits in order to
improve the angular resolution \cite{argo-rec}.

\section{Monte Carlo Simulation}

The air showers development in the atmosphere has been generated
with the CORSIKA v. 6.500 code including the QGSJET-II.03 hadronic
interaction model for primary energy above 80 GeV and the FLUKA
code for lower energies \cite{corsika}. Cosmic ray spectra of p,
He and CNO have been simulated in the energy range from 30 GeV to
1 PeV following \cite{wiebel-sooth}. The relative fractions (in \%
of the total) after triggering by the ARGO-YBJ detector for events
with N$_{strip}\geq$30 are: p$\sim$88\%, He$\sim$10\%,
CNO$\sim$2\%. About 3$\cdot$10$^{11}$ showers have been
distributed in the zenith angle interval 0-60 degrees. The
secondary particles have been propagated down to a cut-off energy
of 1 MeV. The experimental conditions have been reproduced via a
GEANT3-based code. The shower core positions have been randomly
sampled in an energy-dependent area up to 10$^3\times$10$^3$
m$^2$, centered on the detector. Simulated events have been
generated in the same format used for the experimental data and
they have been analyzed by using the same reconstruction code.

A detailed Monte Carlo simulation of cosmic rays propagation in
the Earth-Moon system has been developed to compare the observed
displacement of the shadow with the expectations, so disentangling
the effect of the geomagnetic field from a possible systematic
pointing error. The algorithm is described in \cite{mcmoon}.

\section{Data analysis}

For the analysis of the shadowing effect a 10$^{\circ}\times$
10$^{\circ}$ sky map in celestial coordinates (right ascension and
declination) with 0.1$^{\circ}\times$0.1$^{\circ}$ bin size,
centered on the Moon location, is filled with the detected events.
The background is evaluated with the so-called \emph{time
swapping} method as described in \cite{moon-icrc09}.

To form conservative estimate of the $\bar{p}/p$ ratio in this
preliminary work the analysis refers to about 6$\cdot$10$^8$
events collected only during the stable operation period of data
taking after the following event selection: (1) each event should
fire more than 30 strips on the ARGO-130 central carpet to avoid
any threshold effect; (2) the zenith angle of the event arrival
direction should be less than 50$^{\circ}$; (3) the reconstructed
core position should be inside an area 250$\times$250 m$^2$ around
the detector; (4) the reduced $\chi^2$ of the final temporal fit
should be less than 100 ns$^2$.

 \begin{figure}[!t]
  \centering
  \includegraphics[width=3.0in]{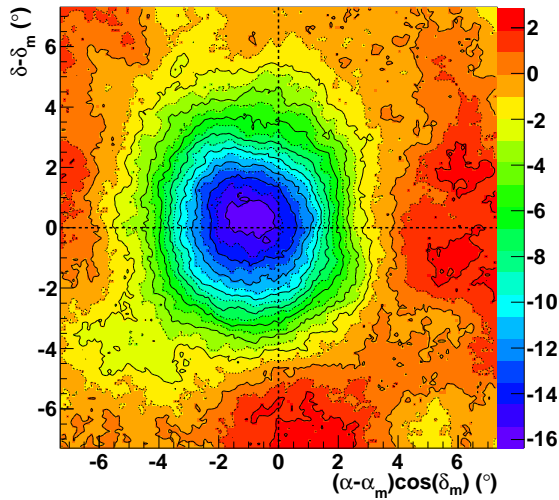}
  \caption{Moon shadow significance map observed by the ARGO-YBJ
detector in 1350 hours on-source for events with N$_{strip}$ = 30
- 60 and zenith angle $\theta<50^{\circ}$. The color scale gives
the statistical significance.}
  \label{fig:moon1}
 \end{figure}

A significance map of the Moon shadow is shown in Fig.
\ref{fig:moon1}. It contains all the events belonging to the
lowest multiplicity bin investigated (N$_{strip}$ = 30 - 60),
collected by ARGO-YBJ during the period December 2007 - December
2008 (1350 hours on-source). The sensitivity of the observation is
about 17 standard deviations and the Moon is shifted westward by
about 1$^{\circ}$. This means that a potential antiproton signal
is expected eastward within 1.5$^{\circ}$ from the actual Moon
position. The median energy of selected events is
E$_{50}\approx$1.4 TeV (mode energy $\sim$ 0.30 TeV) for
proton-induced showers. The corresponding angular resolution is
$\sim$1.8$^{\circ}$ (a detailed analysis of the Moon shadowing
effect is given in \cite{moon-icrc09}).
%

\begin{figure}[!t]
  \centering
  \includegraphics[width=3.0in]{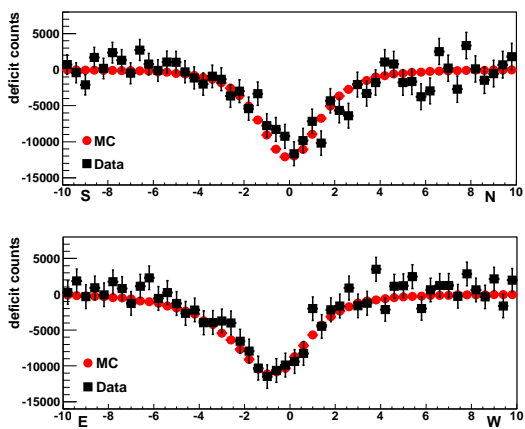}
  \caption{Upper (lower) panel: deficit counts observed around the Moon projected along
the North-South (East-West) axis for N$_{strip}$ = 30 - 60 (black
squares) compared to the MC simulation expectations.}
  \label{fig:proj-3060}
 \end{figure}

The projection along the North-South (East-West) direction of the
deficit counts around the Moon is shown in the upper (lower) panel
of Fig. \ref{fig:proj-3060} for N$_{strip}$ = 30 - 60. The
vertical axis reports the events contained in an angular slice
parallel to the North-South (East-West) axis and centered to the
observed Moon position. The width of this band is
$\pm$3.3$^{\circ}$. The data are in good agreement with the MC
simulation and the observed shadow is shifted westward by about
1$^{\circ}$ (lower panel), as expected. Since at the Yangbajing
latitude the effect of the geomagnetic field along the North-South
direction is negligible, a nearly symmetrical and centered
projection is expected. A detailed analysis of this sort of
projections and of the energy calibration of the detector is given
in \cite{moon-icrc09}. We stress that the systematic pointing
error has been taken into account in the upper limit calculations.


\begin{figure}[!t]
  \centering
  \includegraphics[width=3.0in]{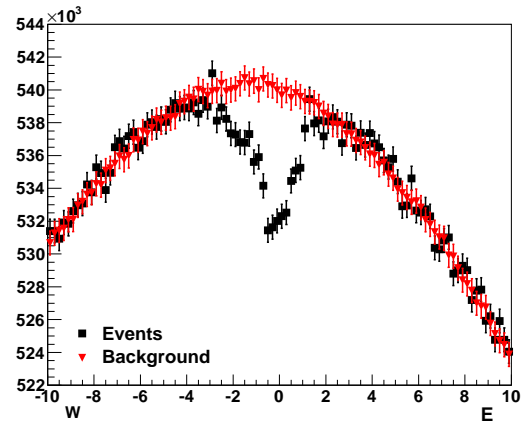}
  \caption{The event and background maps around the Moon projected along
the East-West axis for N$_{strip}>$60 are compared.}
  \label{fig:moon-evbkg}
 \end{figure}

The event and background maps around the position of the Moon
projected along the East-West axis for N$_{strip}>$60 are compared
in Fig. \ref{fig:moon-evbkg}. As can be seen, the shadow clearly
appears in the event map, without any background subtraction.

\section{Results and discussion}

In order to evaluate the $\bar{p}/p$ ratio, a maximum likelihood
fit is performed using the $\bar{p}$ content as a free parameter
with following procedure:
\begin{enumerate}
 \item the Moon shadow EW projection has been drawn both for data and MC.
 \item the MC Moon shadow has been split into a "matter" part
 \emph{plus} an "antiproton" part, in such a way that the
 \emph{total amount of triggered events remains unchanged}:
  \setlength{\arraycolsep}{0.0em}
  \begin{eqnarray*}
\Phi_{MC}(mat) \longrightarrow \Phi_{MC}(r;mat+\bar{p})\nonumber\\
= (1-r)\Phi_{MC}(mat)+r\Phi_{MC}(\bar{p})
  \end{eqnarray*}
  \setlength{\arraycolsep}{5pt}
 \item for each matter-to-antiproton ratio, the expected Moon shadow
EW projection $\Phi_{MC}(r;mat+\bar{p})$ is compared with the
experimental one via the calculation of the likelihood function:
\begin{equation}
 log\mathcal L (r)=\sum_{i=1}^B N_i ln[E_i(r)]-E_i(r)-ln(N_i!)
 \label{eq:Likelihood}
\end{equation}
where the N$_i$ is the number of experimental events included
within the $i-\textrm{th}$ bin, while $E_i(r)$ is the number of
events expected within the same bin. The number $E_i(r)$ is
calculated by adding the contribution expected from MC
 ($\Phi_{MC}(r;mat+\bar{p})$) to the \emph{measured} background.
\end{enumerate}

\begin{figure}[!t]
  \centering
  \includegraphics[width=3.0in]{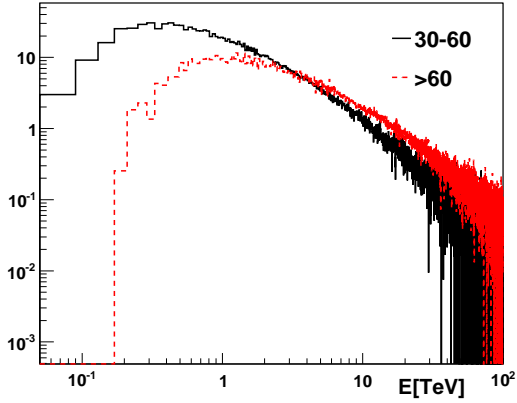}
 \caption{Energy distributions of cosmic ray contributing to the
multiplicity bins: N$_{strip}$ = 30 - 60 and N$_{strip}>$60.}
  \label{fig:energy}
 \end{figure}

This calculation has been performed for two different multiplicity
bins: N$_{strip}$ = 30 - 60 and N$_{strip}>$60. The distributions
of cosmic ray primary energy contributing to these intervals are
shown in Fig. \ref{fig:energy}. From the MC simulation it results
that the median energy of all selected events is E$_{50}$ = 1.85
TeV for N$_{strip}$ = 30 - 60 and E$_{50}$ = 4.10 TeV for
N$_{strip}>$60.

\begin{figure}[!t]
  \centering
  \includegraphics[width=3.0in]{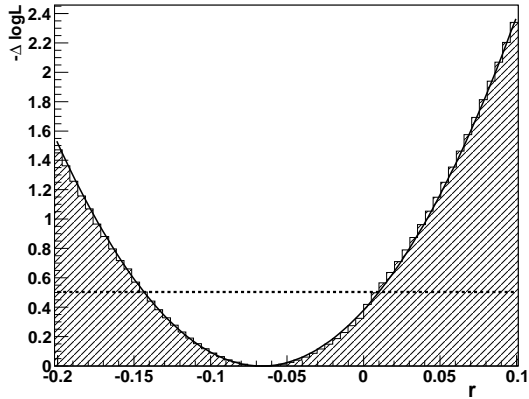}
 \caption{$\Delta log\mathcal L$ as a function of $r$ (the
$\bar(p)$ content) for N$_{strip}$ = 30 - 60. The dashed line
$\Delta log\mathcal L$ = 0.5 is used to determine the 68.3\%
central confidence level.}
  \label{fig:llf}
 \end{figure}

The result of this analysis is shown in Fig. \ref{fig:llf} for
N$_{strip}$ = 30 - 60 and one finds $r_{min}=\Phi (\bar{p})/\Phi
(mat)= -0.065\pm 0.078$. The result is below a physical boundary
(the $\bar{p}$ content must be positive). An upper limit of 7.4\%
with 90\% confidence level is set using the unified approach of
Feldmann \& Cousin \cite{feldman} for N$_{strip}$ = 30 - 60. For
higher multiplicities (N$_{strip}>$60) an upper limit of 7.4\% is
set with 90\% c.l.. Notice that the two values are similar, in
spite of the difference of the multiplicity interval. It is a
consequence of the combination of the two opposing effects of the
angular resolution and of the geomagnetic deviation about which we
said before.
 \begin{figure}[!t]
  \centering
  \includegraphics[width=3.0in]{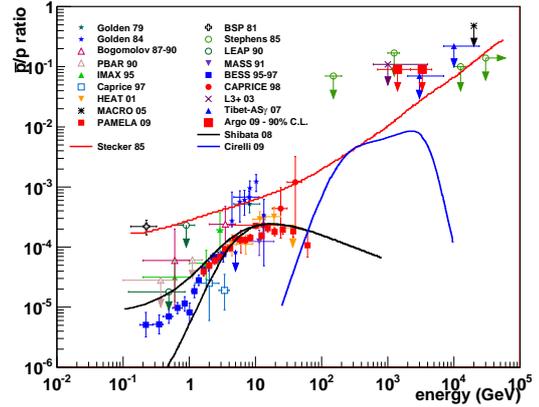}
\caption{The antiproton to proton flux ratio obtained with the
ARGO-YBJ experiment compared with all available measurements. The
solid curve refer to a direct production model. The dashed line
refers to a model of primary $\bar{p}$ production by antistars
\cite{golden}, while the dotted one refers to the contribution of
an heavy DM particle \cite{cirelli}.}
  \label{fig:antipp}
 \end{figure}
With the assumed flux composition in the few-TeV energy range of
88\% protons and 12\% heavier nuclei responsible of the observed
deficit, these limits corresponds to a $\bar{p}/p$ ratios of 0.09
for N$_{strip}$ = 30 - 60 with 90\% c.l. and 0.09 for N$_{strip}>$
60 with 90\% c.l., respectively. In Fig. \ref{fig:antipp} the
ARGO-YBJ results are shown with a compilation of available
measurements \cite{tibet-antip}.

\section{Conclusions}

The ARGO-YBJ experiment, in stable data taking since November 2007
with an energy threshold of a few hundreds of GeV, is observing
the Moon shadow with high statistical significance. Using about 1
year data, a preliminary upper limit of the $\bar{p}/p$ flux ratio is set to
0.09 with a confidence level of 90\% at a median energy of 1.85
and 4.10 TeV. We stress that in the few-TeV range this result is
among the lowest available. In 3 years of data taking the ARGO-YBJ
experiment will be able to lower this limit down to the percent
level, thus excluding some of the current antiproton direct
production models.

\end{document}